\documentclass[12pt]{article}
\usepackage{amsmath}
\usepackage{mathrsfs}
\usepackage{graphicx}
\usepackage{ amssymb }
\usepackage{multirow}
\usepackage[margin=2cm]{geometry} 
\usepackage{slashed}
\usepackage{caption}
\newcommand{\overbar}[1]{\mkern1.5mu\overline{\mkern-1.5mu#1\mkern-1.5mu}\mkern 1.5mu}

\def\un#1{\relax\ifmmode\@@underline#1\else
        $\@@underline{\hbox{#1}}$\relax\fi}

\let\du=\du                     

\def\a{\alpha}
\def\b{\beta}

\def\j{\psi}

\def\m{\mu}
\def\n{\nu}

\def\s{\sigma}

\def\x{\xi}

\def\bo{{\raise-.3ex\hbox{\large$\Box$}}}               
\def\pa{\partial}                                       
\def\TH{{\raise.2ex\hbox{$\displaystyle \bigodot$}\mskip-4.7mu \llap H \;}}
\def\face{{\raise.2ex\hbox{$\displaystyle \bigodot$}\mskip-2.2mu \llap {$\ddot
        \smile$}}}                                      
\def\dg{\sp\dagger}                                     

\def\sp#1{{}^{#1}}                              
\def\VEV#1{\left\langle #1\right\rangle}        
\def\leftrightarrowfill{$\mathsurround=0pt \mathord\leftarrow \mkern-6mu
        \cleaders\hbox{$\mkern-2mu \mathord- \mkern-2mu$}\hfill
        \mkern-6mu \mathord\rightarrow$}
\def\dvec#1{\vbox{\ialign{##\crcr
        \leftrightarrowfill\crcr\noalign{\kern-1pt\nointerlineskip}
        $\hfil\displaystyle{#1}\hfil$\crcr}}}           
\def\dt#1{{\buildrel {\hbox{\LARGE .}} \over {#1}}}     

\newskip\humongous \humongous=0pt plus 1000pt minus 1000pt

\newif\ifdtup

 \newcommand{\be}{\begin{equation}}
\newcommand{\ee}{\end{equation}}
\newcommand{\nbe}{\begin{equation*}}
\newcommand{\nee}{\end{equation*}}

\newcommand{\fr}{\frac}
\newcommand{\lb}{\label}

\def\fracmm#1#2{{{#1}\over{#2}}}
  
\def\low#1{{\raise -3pt\hbox{${\hskip 0.75pt}\!_{#1}$}}}

\begin{document}

\thispagestyle{empty}

{\hbox to\hsize{
\vbox{\noindent July 2018 \hfill IPMU18-0104 }}
\noindent  \hfill }

\noindent
\vskip2.0cm
\begin{center}

{\Large\bf General couplings of a vector multiplet in N=1 supergravity with new FI terms
}

\vglue.3in

Yermek Aldabergenov~${}^{a,b}$, Sergei V. Ketov~${}^{a,b,c}$ and Rob Knoops~${}^{d}$
\vglue.1in

${}^a$~Department of Physics, Tokyo Metropolitan University, \\
Minami-ohsawa 1-1, Hachioji-shi, Tokyo 192-0397, Japan \\
${}^b$~Research School of High-Energy Physics, Tomsk Polytechnic University,\\
2a Lenin Avenue, Tomsk 634050, Russian Federation \\
${}^c$~Kavli Institute for the Physics and Mathematics of the Universe (IPMU),
\\The University of Tokyo, Chiba 277-8568, Japan \\
${}^d$~Department of Physics, Faculty of Science, Chulalongkorn University,
Phayathai Road, Pathumwan, Bangkok 10330, Thailand 
\vglue.1in
aldabergenov-yermek@ed.tmu.ac.jp, ketov@tmu.ac.jp, rob.k@chula.ac.th
\end{center}

\vglue.3in

\begin{center}
{\Large\bf Abstract}
\end{center}
\vglue.1in
\noindent  We propose new interactions of a (massive) vector multiplet with chiral multiplets and (D-type)  spontaneously broken supersymmetry in four-dimensional $N=1$ supergravity, due to the generalized Fayet-Iliopoulos (FI) terms. Our actions are invariant under linearly realized off-shell supersymmetry and K\"ahler-Weyl transformations. We compute the scalar potentials and pinpoint some physical restrictions arising in this approach. 
\newpage

\section{Introduction}

Local and linearly realized supersymmetry significantly restricts the structure of interactions 
in four-dimensional $N=1$ supergravity and imposes severe constraints on its phenomenological applications in particle physics and cosmology, including inflationary models, in particular \cite{Ketov:2012yz}. It is, therefore, of interest to extend the standard framework of $N=1$ supergravity, in order to allow more interactions. One of the known and popular ways in this direction is the use of non-linear realizations of supersymmetry and constrained (nilpotent) superfields 
\cite{Lindstrom:1979kq,Samuel:1982uh,Antoniadis:2014oya,Ferrara:2014kva,Dudas:2015eha}. 
This is fully legitimate from the viewpoint of considering supergravity as the effective theory arising in the low-energy approximation from a more fundamental theory, say, superstrings. However, having a linearly realized supersymmetry is advantageous because it is preserved by quantum corrections. Moreover,  the non-linear realizations of supersymmetry with the constrained superfields can often be reformulated (or derived) by using linear (and manifest) realizations of supersymmetry when the latter is spontaneously broken.  In \cite{Farakos:2018sgq} these ideas were applied to a derivation of new interactions of chiral superfields  in $N=1$ supergravity, by using the new {\it Fayet-Iliopoulos} (FI) term proposed in \cite{Cribiori:2017laj}. The original FI term \cite{Cribiori:2017laj} was found to be violating 
K\"ahler-Weyl invariance in the presence of (chiral) matter.

In this Letter we propose more general FI terms for a construction of new interactions of a (massive)
vector superfield with chiral superfields in $N=1$ supergravity. Our FI terms include arbitrary functions, and  the proposed actions  are invariant under {\it linearly} realized (manifest)  $N=1$ local supersymmetry {\it and\/} K\"ahler-Weyl (gauge) transformations. Supersymmetry is spontaneously broken by a $D$-term, while it is the condition for consistency of new interactions. We compute the scalar potentials in our models, and find some  physical restrictions arising in this approach. The massive vector supermultiplet has a real scalar amongst its
field components, while this scalar is identified with inflaton in some viable models of cosmological inflation and dark matter, based on supergravity \cite{Farakos:2013cqa,Ferrara:2013rsa,Aldabergenov:2016dcu,Aldabergenov:2017bjt,Addazi:2017ulg}.

\section{A vector multiplet in N=1 supergravity}

In this Section we collect the known facts about an $N=1$ vector multiplet coupled to chiral matter in $N=1$ supergravity, by using curved superspace of the old-minimal N=1 supergravity. This section serves as our setup and introduces our notation, as in Ref.~\cite{Wess:1992cp}, with the spacetime signature $(-,+,+,+)$.~\footnote{The reduced Planck mass $M_P$ is set to one for simplicity.} 

The standard general Lagrangian reads
\begin{equation} \label{lag}
\mathcal{L}=-3\int d^4\theta E e^{-\hat{K}/3} + \left( \int d^2\Theta 2\mathcal{E}\left[  \mathcal{W} +\frac{1}{4}f W^\alpha W_\alpha\right] +h.c.\right)~,
\end{equation}
in terms of a K\"ahler potential $K(\Phi,\bar{\Phi},H,\bar{H},V)$ depending upon neutral chiral superfields  $(\Phi,\bar{\Phi})$ and charged
chiral superfields $(H,\bar{H})$ coupled to a general vector gauge superfield $V$, subject to the supergauge transformations
\begin{equation} \label{sgtr}
H\rightarrow e^{-iZ}H~,\quad \overbar{H}\rightarrow e^{i\overbar{Z}}\overbar{H}~,\quad 
V\rightarrow V+\frac{i}{2}(Z-\overbar{Z})~,
\end{equation}
with the chiral superfield gauge parameter $Z$. The K\"ahler potential is gauge-invariant provided that the gauge superfield $V$ enters it via the combination $\bar{H}e^{2V}H$. The superpotential $\mathcal{W}(\Phi,H)$ and the kinetic function $f(\Phi,H)$ are gauge-invariant also. The gauge-invariant vector superfield strength is defined by 
\begin{equation} \label{gss}
W_{\a} = -\fracmm{1}{4}  \left( \bar{\mathcal D}^2 - 8{\mathcal R}\right){\mathcal D}_{\a}V~,
\end{equation}
where $\mathcal R$ is the chiral curvature superfield. The $W_{\a}$ obeys Bianchi identities
\begin{equation} \label{bis}
\bar{\mathcal D}_{\dt{\b}}W\low{\a}=0 \quad {\rm and} \quad \bar{\mathcal D}_{\dt{\a}}\bar{W}^{\dt{\a}}
\equiv \bar{{\mathcal D}}\bar{W}={\mathcal D}^{\a}W_{\a}\equiv {\mathcal D}W~.
\end{equation}

The Lagrangian (\ref{lag}) is also invariant under K\"ahler gauge transformations 
\be \lb{kwt}
\hat{K}\rightarrow \hat{K} + 6\Sigma +6\bar{\Sigma}~,\quad  \mathcal{W}\rightarrow  e^{-6\Sigma}\mathcal{W}~,
\quad W_{\a} \rightarrow e^{-3\Sigma}W_{\a}~,
\ee
with the chiral superfield parameter $\Sigma$, accompanied by Weyl transformations of the superspace densities \cite{Howe:1978km},
\be 
d^4\theta E\rightarrow d^4\theta E e^{ 2\Sigma+2\bar{\Sigma}} \quad {\rm and} \quad 
d^2\Theta 2\mathcal{E} \rightarrow d^2\Theta 2\mathcal{E}e^{6\Sigma}~.
\ee
The gauge superfield $V$ is inert under these transformations, whereas the other relevant quantities transform as \cite{Howe:1978km}
\be \lb{kt2}
\mathcal D_\alpha V  \rightarrow e^{\Sigma - 2 \bar \Sigma}\mathcal D_\alpha V~, \quad
(\bar {\mathcal D}^2 - 8 \mathcal R)  \rightarrow 
(\bar {\mathcal D}^2 - 8 \mathcal R)  e^{-4 \Sigma + 2 \bar\Sigma}~, \quad
\mathcal DW \rightarrow e^{- 2 \Sigma - 2 \bar \Sigma} \mathcal D W~.
\ee

The field components of a superfield are identified with the leading fields of its superspace covariant derivatives as  $\left.\Phi\right| = \phi$, $\left.V\right| = C$, etc.  In particular, the vector superfield strength components are defined by
\be \lb{vscomp}
\left. W_{\a} \right| =\j_{\a}~,\quad \left. {\mathcal D}W\right|=-2D \quad {\rm and} \quad \left. {\mathcal D}_{(\a}W_{\b)}\right|=
i\s^{ab}_{\a\b}\hat{F}_{ab}=2i\hat{F}_{\a\b}~,
\ee
where $\j_{a}$ is Majorana fermion (photino), $D$ is the real auxiliary field, $\hat{F}_{ab}$ is the (supercovariantized) abelian vector field
strength, $\hat{F}_{ab}=\pa_aV_b-\pa_bV_a +{\rm fermionic~terms}$. It also implies
\be \lb{d2w2}
\left. -\frac{1}{4}{\mathcal D}^2W^2\right|=D^2 -2F^{\a\b}F_{\a\b} + {\rm fermionic~terms}~.
\ee

It is worth mentioning that our setup can be consistently elevated to the conformal supergravity framework, either in curved superspace or in superconformal tensor calculus \cite{Freedman:2012zz}, by using compensators. Actually, we used both approaches to control our calculations.

\section{Spontaneous SUSY breaking by FI terms}

 It is possible to add more supersymmetric matter couplings, as well as vector multiplet self-couplings, when supersymmetry is spontaneously broken by  
\be \lb{dvev}
\VEV{D}\equiv \x \neq 0~.
\ee
The first such self-coupling (with linearly realised local supersymmetry)  proposed in \cite{Cribiori:2017laj} in the case of a single vector multiplet reads
\begin{equation} \label{cri}
\mathcal{L}_{\x}=8\xi\int d^4\theta E\frac{W^2\overbar{W}^2}{\mathcal{D}^2W^2\overbar{\mathcal{D}}^2\overbar{W}^2}\mathcal{D}W~.
\end{equation}
It can be interpreted as a {\it Fayet-Iliopoulos} (FI) term \cite{Fayet:1974jb}, because in the absence of matter fields it gives rise to a linear term proportional to $\xi D$ in the scalar potential, and hence, to (\ref{dvev}) as well. It is worth mentioning that the  FI term (\ref{cri}) does not require gauging the R-symmetry, and is deeply related to non-linear realizations of supersymmetry and constrained (nilpotent) superfields \cite{Cribiori:2017laj}, unlike the  standard FI term in supergravity 
\cite{Freedman:1976uk,Binetruy:2004hh}. 

However, in the presence of chiral matter,  the FI term (\ref{cri}) is not invariant under the K\"ahler-Weyl transformations, and the factors $\mathcal{D}^2W^2$ or $(\mathcal{D}^2-8{\mathcal R^{\dg}})W^2$ do
not transform covaraintly. Also, the bosonic contribution to the scalar potential (in Einstein frame) reads
\be \lb{xb}
e^{-1}{\mathcal L}_{FI}=-\x e^{\hat{K}/3}D,
\ee
so that the $D$-equation of motion yields the scalar potential contribution 
\be \lb{xcpot}
V_{FI} = \fr{1}{2} \x^2 e^{2\hat{K}/3}
\ee
that is obviously not K\"ahler gauge invariant. It was proposed  in \cite{Antoniadis:2018oeh} to cure this problem by inserting the factor of $e^{-\hat{K}/3}$ into the FI term (\ref{cri}), in order to compensate the $e^{\hat{K}/3}$ factor in (\ref{xb}). In Einstein frame, it gives rise to the simplest FI term $-\x D$ with the simplest (field-independent) contribution $\fr{1}{2} \x^2$ to the scalar potential  \cite{Antoniadis:2018oeh}. Though this is enough for a spontaneous SUSY breaking and uplifting of an AdS or a Minkowski vacuum, if any,  to a dS vacuum, it is not enough for viable phenomenological applications, because it identifies  the SUSY breaking scale with the dark energy scale (described by the cosmological constant).

We propose to generalize the FI term (\ref{cri}) further, by promoting a FI constant $\x$ to a  K\"ahler- and gauge- invariant  "superfield-dependent" FI coupling $\x'$ and inserting the "compensating" powers of 
$e^{\hat K}$ (in part, forming the invariant combination $(\mathcal{D}^2-8{\mathcal R}^\dagger) W^2e^{2\hat{K}/3}$) as follows:~\footnote{In the superconformal approach \cite{Cribiori:2017laj}, this corresponds to 
 \begin{align}  \tilde {\mathcal L}_1 &= - \left[ \mathcal R \   (V)_D \  \left( \xi_1 + U_1 (\Phi, \bar \Phi, H, \bar H, V) \right)    \right]_D \nonumber \end{align}
with
 \begin{align} \mathcal R = (S_0 \bar S_0 e^{-\hat{K}/3} )^{-3}  \frac{(\bar \lambda P_R \lambda) (\bar \lambda P_L \lambda)}  { T \left( \frac{\lambda P_R \lambda}{(S_0 \bar S_0 e^{-\hat{K}/3})^2 } \right) 	\bar T \left( \frac{\lambda P_L \lambda}{(S_0 \bar S_0 e^{-\hat{K}/3})^2    }\right)  }~~. \nonumber \end{align}
}
\begin{equation} \label{our1}
\mathcal{L}_{1}=8\int d^4\theta E \frac{W^2\overbar{W}^2}{\left[ (\mathcal{D}^2-8{\mathcal R}^\dagger) W^2e^{2\hat{K}/3}\right] \left[(\bar{\mathcal D}^2-8{\mathcal R)}\overbar{W}^2e^{2\hat{K}/3} \right]}(\mathcal{D}W) e^{\hat{K}}   \x_1'(\Phi,\bar{\Phi},H,\bar{H},V)~.
\end{equation}
It is convenient to extract from $\x_1'(\Phi,\bar{\Phi},H,\bar{H},V)$ its (constant) vacuum expectation value,
\be \lb{xvev}
\xi_1'(\Phi, \bar \Phi, H, \bar H, V)=\x_1 + U_1(\Phi, \bar \Phi, H, \bar H, V)\quad {\rm with}\quad  \VEV{U_1}=0 \quad
{\rm and} \quad \x_1\neq 0~.
\ee

Different FI terms were proposed in \cite{Kuzenko:2018jlz}. We select the simplest one of them, having the form
\begin{equation} \label{our2}
\mathcal{L}_{2}=8\int d^4\theta E\frac{W^2\overbar{W}^2}{(\mathcal{D}W)^2(\overbar{\mathcal D}\overbar{W})^2}  (\mathcal{D}W) e^{-\hat{K}/3}  \left(\x_2 +U_2(\Phi,\bar{\Phi},H,\bar{H},V)\right)~,
\end{equation}
where we have inserted the factor of $e^{-\hat{K}/3}$, in order to maintain the K\"ahler-Weyl invariance, and have added an invariant function $U_2(\Phi,\bar{\Phi},H,\bar{H},V)$ to the FI constant $\x_2$. ~\footnote{Its superconformal form
reads
\begin{align}
 \tilde{\mathcal L}_{2} = - \frac{1}{4} \left[ S_0 e^{-\hat{K}/3} \bar S_0  \frac{(\bar \lambda P_L \lambda) (\lambda P_R \bar \lambda)}{ \left( (V)_D \right)^3 } 
 \left( \xi_2 + U_2(\Phi, \bar \Phi, H, \bar H, V) \right) \right]_D~~. \nonumber
\end{align}
}

\section{Scalar potential and kinetic terms}

In this Section we add one of the generalized FI terms (\ref{our1}) or (\ref{our2}) to the standard Lagrangian (\ref{lag}),
and investigate the resulting  new supergravity theory in components.~\footnote{It is also possible to
add both FI terms. However, this does not add new qualitative features.}

As regards the $D$-type scalar potential $\mathcal{V}_D$, the $D$-dependent contributions to the Lagrangian are given by a sum of the standard quadratic term and a linear term only. We find (in Einstein frame, with a single $H$)
\be \lb{spot}
\mathcal{V}_D = \fr{1}{2} ({\rm Re}f)^{-1}\left(\x + U -\fr{1}{2}\hat{K}_V\right)^2
\ee
for each of the FI terms (\ref{our1}) and (\ref{our2}), where we have introduced the notation 
$\hat{K}_V=\left. \fr{\pa \hat{K}}{\pa V}\right|$. This extends the old result \cite{VanProeyen:1979ks} to the case of a single massive vector multiplet in the presence of chiral matter superfields with a generalized FI term.

Equation (\ref{spot}) can be further generalized to the case of several charged chiral superfields $H^{m}$ with the leading 
field components  $h^{m}$, when the K\"ahler potential $\hat{K}$ allows Killing vectors $k^{m}$, and the superpotential $\mathcal W$ has the property
\be \lb{hom} 
k^m \fr{\pa \mathcal W}{\pa H^{m}} = - r\mathcal{W}.
\ee
This is the case when, for example, the superpotential transforms under the gauge symmetry as ${\mathcal W} \rightarrow \mathcal{W} e^{- i \xi \Lambda}$, and the chiral superfields transform as $H \rightarrow H e^{-i \Lambda}$, while its simplest realization is $\mathcal{W} \propto  H^\xi$ ---  then we have $r = i \xi$, where $\xi$ is the FI constant. Under these conditions, the scalar potential  is again given by (\ref{spot}) after a substitution of $-\fr{1}{2}\hat{K}_V$ by the moment map $\mathcal P$ defined by 
\be \lb{mp}
\mathcal P = i (k^m \partial_m \hat{K} - r )
\ee
that is both gauge- and K\"ahler- invariant. It is worth noticing that $r\neq 0$ is also relevant in the
case of the gauged R-symmetry by the use of the gauge superfield $V$.

We also find that in the case of FI term (\ref{our1}), there are {\it no} purely bosonic interactions depending upon $F_{\m\n}$
from the FI term alone.  However, in addition to the condition (\ref{dvev}), its fermionic terms are well defined only if their
denominators  do not hit {\it singularities} when either $(F^-)^2$ or $(F^+)^2$ approach $\VEV{D}^2=\x^2$.~\footnote{We
define the (anti)self-dual tensors as $F_{\mu \nu}^\pm = \frac{1}{2} ( F_{\mu \nu} \pm \tilde F_{\mu \nu})$, $F_{\mu \nu}^\pm = (F_{\mu \nu}^\mp)^*$ and $\tilde F^{\mu \nu} = - \frac{i}{2}  \epsilon^{\mu \nu \rho \sigma} F_{\rho \sigma}$.} This observation put the leverage from above on possible values of the electric field component of $F_{\mu \nu}$ as
\be \lb{elb}  \fr{1}{2}\vec{E}^2 < \xi^2~. \ee 

In the case of FI term (\ref{our2}), the factors $(D^2-(F^{\pm})^2)$ do not appear in denominators of the fermionic terms, 
but arise in the numerator of the bosonic terms as follows:
\begin{align}
e^{-1} \mathcal L_{\rm FI-bos.} 
 &=  - \xi_2'  \frac{1}{D^3} \left( D^2 - \hat F^- \cdot \hat F^- \right) \left( D^2 - \hat F^+ \cdot \hat F^+  \right), \label{xi2_contr}
\end{align}
where we have
\begin{align}  \xi_2' (\phi, \bar \phi, h, \bar h, C) = \xi_2 + U_2 ( \phi, \bar \phi, h, \bar h, C )~, \end{align}
and the covariant field strength $\hat F$ can be found in eq.~(17.1) of Ref.~\cite{Freedman:2012zz}.

Equation (\ref{xi2_contr}) contributes to the kinetic (quadratic in $F$) bosonic terms as 
\begin{align}
e^{-1} \mathcal L_{\rm FI-bos.} &= - \frac{\xi_2'}{D^3} \left( D^2 - \frac{1}{4} \left(F \cdot F + \tilde F \cdot \tilde F \right) \right)^2 \notag \\
&= - \xi_2' D + \frac{\xi_2'}{2D} \left(F \cdot F + \tilde F \cdot \tilde F \right) + \mathcal O (F^4)~.
\end{align}
Having restricted ourselves to the quadratic terms with respect to $F_{\mu \nu}$ and neglecting the higher order terms 
 in the total Lagrangian, we arrive at
\begin{align}
e^{-1}\mathcal L= & ~~\frac{1}{2} (\text{Re}f) D^2 - (\mathcal P + \xi_2') D + \frac{\xi_2'}{D} F \cdot F  \notag \\
&- \frac{1}{4} (\text{Re}f) F\cdot F + \frac{1}{8} ({\rm Im}f)\epsilon^{\mu \nu \rho \sigma} F_{\mu \nu}F_{\rho \sigma} ~~.
\label{xi_2_Full}
\end{align}
where we have used the identity $\tilde F \cdot \tilde F = F \cdot F$.

In order to solve the algebraic equation of motion for $D$, having the form
\begin{align}
- (\mathcal P + \xi_2') + (\text{Re}f) D - \frac{\xi_2'}{D^2} F \cdot F  = 0,
\end{align}
we search for its solution as
\begin{align}
\langle D \rangle = D_0 + Y(\phi, \bar \phi, h, \bar h) F \cdot F 
\end{align}
and find
\be
D_0 = \frac{1}{\text{Re}f} \left( \mathcal P + \xi_2' \right)~, \quad 
Y = \xi_2' \frac{\text{Re}f}{(\mathcal P + \xi_2')^2}~~.
\ee

The bosonic contribution to the Lagrangian under investigation thus reads
\begin{align}
e^{-1}\mathcal L &= 
\left( -\frac{1}{4} +  \frac{\xi_2'}{\mathcal P + \xi_2'} \right) (\text{Re}f) F_{\mu \nu}F^{\mu \nu}
- \mathcal V_D
\end{align}
with the scalar potential
\begin{align}
\mathcal V_D &= \frac{1}{2} \frac{1}{\text{Re}f} \left( \mathcal P + \xi_2' \right)^2 .
\end{align}

The kinetic term of $F_{\mu \nu}$ has the physical sign (no ghosts) when 
\begin{align} \lb{bound}
\frac{\xi_2'}{\mathcal P + \xi_2'} < \frac{1}{4}~.
\end{align}
This condition is violated when either $\mathcal P>3\x_2' $ (provided $\mathcal P + \xi_2' >0$) or
$\mathcal P<3\x_2' $ (provided $\mathcal P + \xi_2' <0$), as well as when $\mathcal P=0$, and
thus excludes the FI term (\ref{our2}) in all these cases. However, this restriction can be easily removed by using a superpotential that leads to non-vanishing vacuum expectation values of physical scalars of the chiral superfields contributing to supersymmetry breaking.~\footnote{When $\mathcal P=0$, adding both FI terms (\ref{our1}) and
(\ref{our2}) obeys the no-ghost condition provided that $\x_2/(\x_2+\x_1) <1/4$.}

\section{Conclusion}

Our new FI terms (\ref{our1}) and (\ref{our2}) fully respect the symmetries of the original action (\ref{lag}), such as local supersymmetry, the gauge invariance and the K\"ahler-Weyl invariance, while all of these
symmetries are manifest. They also include new arbitrary functions that appear in the scalar potential too. The FI term (\ref{our2}) is apparently simpler than (\ref{our1}).

The total scalar potential in the theory (\ref{lag}) is a sum of the $D$-type and $F$-type terms,
\be \lb{ptot}
\mathcal{V}_{\rm tot.} = \mathcal{V}_{D}+\mathcal{V}_{F}~,
\ee
where $V_F$ is given by the standard expression \cite{Cremmer:1978iv}
\be \lb{fp}
\mathcal{V}_{F} = e^{\hat{K}}\left[ D_{\bar{A}}\bar{\mathcal W}g^{\bar{A}B}D_B{\mathcal W} 
- 3\bar{\mathcal W}{\mathcal W} \right]
\ee
in terms of the derivatives $D_A{\mathcal W}={\mathcal W}_A+\hat{K}_A{\mathcal W}$ and the inverse K\"ahler metric $g^{\bar{A}B}$, where the superfield subscripts denote the derivatives with respect to chiral superfields $\Phi^A$. In \cite{Farakos:2018sgq}, the generalized term similar to (but different from) (\ref{our1}) with an arbitrary function $U_3(\bar{\Phi},\Phi)$ inside was introduced for the chiral superfields $\Phi$ by using their K\"ahler potential $K$ instead of $V$ in the definition of the spinor chiral superfield (\ref{gss}).  The
net effect on the $F$-type scalar potential of the chiral superfields is a shift  
\be \lb{shift}
{\mathcal V}_F \to {\mathcal V}_F + U_3
\ee
that results in a totally arbitrary function in the place of ${\mathcal V}_F$ --- it was dubbed as the "liberated supergravity" in  \cite{Farakos:2018sgq}. In the case of the $D$-type "liberated supergravity" studied here, the liberation by the generalized FI terms  appears to be limited: though the scalar potential (\ref{spot}) also includes an arbitrary  function $U$ at our disposal, it is non-negative for any choice of $U$, being given by an arbitrary real function squared.
\vglue.2in

\section*{Acknowledgements}

YA and SVK were supported in part by the Competitiveness Enhancement Program of Tomsk Polytechnic University in Russia. SVK was also supported in part by a Grant-in-Aid of the Japanese Society for Promotion of Science (JSPS) under No.~26400252,  and the World Premier International Research Center Initiative (WPI Initiative), MEXT, Japan. RK was supported by the CUniverse research promotion project of Chulalongkorn University under the grant reference CUAASC.
\vglue.2in

\bibliographystyle{utphys} 

\providecommand{\href}[2]{#2}\begingroup\raggedright
\endgroup

\end{document}